\documentclass[twocolumn,aps,prl,amsfonts,amsmath,showpacs,superscriptaddress,numbers]{revtex4}
\usepackage[utf8]{inputenc}
\usepackage{graphicx,bm} 
\usepackage{amsmath,amssymb,amsfonts}
\usepackage{epstopdf}
\usepackage{array} 

\DeclareMathOperator{\F}{\mathbf{F}}
\DeclareMathOperator{\Z}{\mathbf{Z}}
\DeclareMathOperator{\E}{\mathbb{E}}
\DeclareMathOperator{\G}{\mathbf{G}}

\DeclareMathOperator{\X}{\mathbf{X}}

\newcommand{\cZ}{\mathcal Z}

\begin{document}



\title{Noise-Induced Synchronization, Desynchronization, and Clustering in Globally Coupled Nonidentical Oscillators}

\author{Yi Ming \surname{Lai}}

\affiliation{Department of Mathematics and Statistics, University of Strathclyde, Glasgow G1 1XP, UK}

\affiliation{Mathematical Institute, University of Oxford, Oxford, OX1 3LB, UK}

\author{Mason A. Porter}

\affiliation{Mathematical Institute, University of Oxford, Oxford, OX1 3LB, UK}

\affiliation{CABDyN Complexity Centre, University of Oxford, Oxford, OX1 1HP, UK}

\date{\today}



\begin{abstract}

We study ensembles of globally coupled, nonidentical phase oscillators subject to correlated noise, and we identify several important factors that cause noise and coupling to synchronize or desychronize a system. By introducing noise in various ways, we find a novel estimate for the onset of synchrony of a system in terms of the coupling strength, noise strength, and width of the frequency distribution of its natural oscillations.  We also demonstrate that noise alone is sufficient to synchronize nonidentical oscillators.
 However, this synchrony depends on the first Fourier mode of a phase-sensitivity function, through which we introduce common noise into the system. We show that higher Fourier modes can cause desychronization due to clustering effects, and that this can reinforce clustering caused by different forms of coupling. Finally, we discuss the effects of noise on an ensemble in which antiferromagnetic coupling causes oscillators to form two clusters in the absence of noise. 

\end{abstract}

\setcounter{page}{1}
\maketitle



\section{Introduction}

Synchronization describes the adjustment of rhythms of self-sustained oscillators due to their interaction \cite{scholarsync}. Such collective behavior has important ramifications in myriad natural and laboratory systems---ranging from conservation and pathogen control in ecology \cite{cite1} to applications throughout physics, chemistry, and engineering \cite{cite2,kuramoto84}. 


Numerous studies have considered the effects of coupling on synchrony using model systems such as Kuramoto oscillators \cite{cite7}. In a variety of real-world systems, including sets of neurons \cite{cite3} and ecological populations \cite{cite4}, it is also possible for synchronization to be induced by noise. In many such applications, one needs to distinguish between extrinsic noise common to all oscillators (which is the subject of this paper)  and intrinsic noise, which affects each oscillator separately. Consequently, studying oscillator synchrony can also give information about the sources of system noise \cite{dunlopetal}. 
Nakao et al. \cite{nakaoetal} recently developed a theoretical framework for noise-induced synchronization using phase reduction and averaging methods on an ensemble of uncoupled identical oscillators. They demonstrated that noise alone is sufficient to synchronize a population of identical limit-cycle oscillators subject to independent noises, and similar ideas have now been applied to a variety of applications \cite{bresslofflai11,kurebayashietal,cite5}.  

Papers such as \cite{nakaoetal,bresslofflai11,kurebayashietal,cite5} characterized a system's synchrony predominantly by considering the probability distribution function (PDF) of phase differences between pairs of oscillators. This can give a good qualitative representation of ensemble dynamics, but it is unclear how to subsequently obtain quantitative measurements of aggregate synchrony \cite{cite6}. It is therefore desirable to devise new order parameters whose properties can be studied analytically (at least for model systems).

Investigations of the combined effects of common noise and coupling have typically taken the form of studying a PDF for a pair of coupled oscillators in a specific application \cite{cite6,cite6b}. Recently, however, Nagai and Kori \cite{nagaikori} considered the effect of a common noise source in a large ensemble of globally coupled, nonidentical oscillators. They derived some analytical results as the number of oscillators $N\rightarrow \infty$ by considering a nonlinear partial differential equation (PDE) describing the density of the oscillators and applying the Ott-Antonsen (OA) ansatz \cite{ottantonsen08,ottantonsen09}. 

In the present paper, we consider the interaction between noise and coupling. We first suppose that each oscillator's natural frequency ($\omega$) is drawn from a unimodal distribution function. For concreteness, we choose a generalized Cauchy distribution 
\begin{equation}
	f_{\mathrm{fr}}(\omega)=\frac{1}{\pi}\frac{\gamma}{\gamma^2+(\omega-\omega_0)^2}\,, 
\end{equation}	
whose width is characterized by the parameter $\gamma$.  The case $\gamma = 1$ yields the Cauchy-Lorentz distribution, and $\omega_0$ is the mean frequency.  We investigate the effects on synchrony of varying the distribution width. Taking the limit $\gamma \rightarrow 0$ yields the case of identical oscillators; by setting the coupling strength to $0$, our setup makes it possible to answer the hitherto unsolved question of whether common noise alone is sufficient to synchronize nonidentical oscillators. 

We then consider noise introduced through a general phase-sensitivity function \footnote{Reference \cite{nagaikori} considered only the function $\sin\theta$.}, which we express in terms of Fourier series. When only the first Fourier mode is present, we obtain good agreement between theory and simulations. However, our method breaks down when higher Fourier modes dominate, as clustering effects \cite{nakaoetal,bresslofflai11} imply that common noise can cause a decrease in our measure of synchrony. Nevertheless, we show that such noise can reinforce clustering caused by different forms of coupling.
Finally, we consider noise-induced synchrony in antiferromagnetically coupled systems, in which pairs of oscillators are negatively coupled to each other when they belong to different families but positively coupled to each other when they belong to the same family.


\section{Globally Coupled Oscillators With Common Noise} 

We start by considering globally coupled phase oscillators subject to a common external force: 
\begin{equation}
	\frac{d\theta_i}{dt}=\omega_i+\frac{K}{N}\sum_{j=1}^N \sin(\theta_j-\theta_i)+\sigma\cZ(\theta_i)p(t)\,, \label{eq:1}
\end{equation}
where $\theta_i$ and $\omega_i$ are (respectively) the phase and natural frequency of the $i$th oscillator, $K>0$ is the coupling strength, $p(t)$ is a common external force, the parameter $\sigma\geq0$ indicates the strength of the noise, and the \emph{phase-sensitivity function} $\cZ(\theta)$ represents how the phase of each oscillator is changed by noise. As in Ref.~\cite{nagaikori}, we will later assume that $p(t)$ is Gaussian white noise, but we treat it as a general time-dependent function for now. 


As mentioned above, $\cZ(\theta)$ indicates how the phase of each oscillator is affect by noise. Such a phase sensitivity function can also be used for deterministic perturbations (e.g., periodic forcing). In the absence of coupling, one can envision that equation (\ref{eq:1}) is a phase-reduced description of an $N$-dimensional dynamical system that exhibits limit-cycle oscillations and which is then perturbed by extrinsic noise: 
\begin{equation}
	\frac{d\X}{dt}=\F(\X)+\sigma \G(\X)p(t)\,.\label{eq:1.5}
\end{equation}
One can reduce (\ref{eq:1.5}) to a phase-oscillator system of the form $\frac{d\theta}{dt}=\omega+\sigma \Z(\X) \cdot \G(\X) p(t)$, where $\Z(\X)$ is the phase resetting curve (PRC) \cite{yoshimuraarai}. In this case, $\cZ(\theta) = \Z(\X) \cdot \G(\X)$. 

We study the distribution of phases $f(\omega,\theta,t)$ in the $N\rightarrow \infty$ limit. First, we define the (complex) Kuramoto order parameter $r(t)=\int_{-\infty}^{\infty}\int_0^{2\pi} \exp(i\theta) f(\omega,\theta,t) d\theta d\omega$. The magnitude $|r|$ characterizes the degree of synchrony in the system, and the phase $\varphi := \arg(r)$ gives the mean phase of the oscillators. From equation (\ref{eq:1}), it then follows that the instantaneous velocity of an oscillator with frequency $\omega$ at position $\theta$ is $\omega+K|r|\sin(\varphi-\theta)+\sigma \cZ p(t)$. Combined with the normalization condition $\int_0^{2\pi} f(\omega,\theta,t)d\theta=1$, the conservation of oscillators of frequency $\omega$ then implies that the phase distribution $f$ satisfies the nonlinear Fokker-Planck equation (FPE)
\begin{equation}
	\frac{\partial f}{\partial t}+\frac{\partial}{\partial \theta}\left[\{\omega+\frac{K}{2i}(r e^{-i\theta}-r^*e^{i\theta})+\sigma \cZ p(t) \}f\right]=0\,. \label{eq:2}
\end{equation}
For more details about the derivation of this evolution equation, see Ref.~\cite{cite7}. To obtain an equation for the order parameter $r$, we follow the approach of Nagai and Kori \cite{nagaikori} and use the OA ansatz that the phase distribution is of the form 
\begin{equation}\label{expand1}
	f = \frac{f_{\mathrm{fr}}(\omega)}{2\pi}\left(1+\sum_{n=1}^\infty\left[(\alpha \exp (i\theta))^n+(\alpha^*\exp(-i\theta))^n\right]\right)\,, 
\end{equation}	
where $\alpha=\alpha(\omega,t)$ is a complex-valued function. This form of $f_{\mathrm{fr}}$ makes it possible to perform contour integration and obtain $r(t)=\alpha^*(\omega_0-\gamma i,t)$. See Ref.~\cite{ottantonsen08} for a discussion about multimodal $f_{\mathrm{fr}}$.


We express the phase-sensitivity function $\cZ$ in terms of its Fourier series: 
\begin{align}\label{expand2}
	\cZ(\theta) &=\frac{a_0}{2}+ \sum_{m=1}^{\infty}[a_m\cos(m\theta)+b_m\sin(m\theta)] \notag \\
		&= c_0+\sum_{m=1}^\infty[c_m \exp(im\theta)+c_m^*\exp(-im\theta)]\,, 
\end{align}
where $c_m=(a_m-ib_m)/2$. We substitute the series expansions (\ref{expand1}) and (\ref{expand2}) into (\ref{eq:2}) to obtain 
\begin{align}
	\frac{dr}{dt}&=-\gamma r+ i\omega_0 r+\frac{K}{2}r\left(1-|r|^2\right) \notag \\ &+ i\sigma p\left(c_0r+c_1^*+\sum_{n=1}^{\infty}{r^*}^n c^*_{n+1}+\sum_{n=2}^{\infty}r^nc_{n-1}\right)\,. \label{eq:4}
\end{align}
To study the magnitude of $r$, we let $r=\sqrt{A}\exp[i(\omega_0 t+\phi)]$, where $A$ and $\phi$ are real. We express the Fourier coeffiicients of $\cZ$ in terms of their real and imaginary parts using $c_m=(a_m - ib_m)/2$ and then collect real and imaginary terms to get
\begin{align}
	\frac{dA}{dt}&=h(A)+\sigma g_A(A,\omega_0 t+\phi)p\,, \label{eq:7} \\
	\frac{d\phi}{dt}&=\sigma g_\phi(A,\omega_0 t+\phi)p\,, \label{eq:8}
\end{align}
where 
\begin{widetext}
\begin{align}
	h(A) &= \,(K-2\gamma)A-KA^2\,, \notag \\ 
g_A(A,\omega_0 t+\phi) &=\, \sum_{n=1}^{\infty}A^{n/2}(1-A)\bigl\{a_{n}\sin (n[\omega_0 t +\phi]) - b_{n}\cos(n[\omega_0 t +\phi])\bigr\}\,, \notag \\
	 g_\phi(A,\omega_0 t+\phi) &= \,\frac{1}{2}a_0+\sum_{n=1}^{\infty} A^{(n-2)/2}(1+A) \left\{a_n\cos(n[\omega_0 t+\phi])+b_n\sin(n[\omega_0 t +\phi])\right\}\,.
\end{align} 
\end{widetext}
Thus far, we have not made any assumptions about the form of the external driving function $p(t)$, but we now set it to be Gaussian white noise. If the correlation times of the noise is comparable to the amplitude relaxation time of a limit-cycle oscillator, then one might need additional terms to describe the exact phase dynamics \cite{cite9}. However, such terms do not affect long-time phase diffusion and synchronization \cite{kurebayashietal}. 


As $A$ and $\phi$ are now stochastic variables, we would like to study their joint PDF $\varrho(A,\phi,t)$. Treating equations (\ref{eq:7}) and (\ref{eq:8}) as It\={o} stochastic differential equations (SDEs) yields an FPE for the temporal evolution of $\varrho(A,\phi,t)$:
\begin{widetext}
\begin{align}
	\frac{\partial \varrho}{\partial t} = -\frac{\partial }{\partial A}\left(h+\frac{\sigma^2}{2}\left[g_A\frac{\partial g_A}{\partial A}+g_{\phi}\frac{\partial g_{\phi}}{\partial \phi}\right]\right)\varrho -\frac{\sigma^2}{2}\frac{\partial}{\partial \phi}\left(g_{\phi}\frac{\partial g_{\phi}}{\partial \phi} +g_{A}\frac{\partial g_{\phi}}{\partial A}\right)\varrho+\frac{\sigma^2}{2}\left(\frac{\partial^2g_A^2\varrho}{\partial A^2}+2\frac{\partial^2g_Ag_{\phi} \varrho}{\partial A\partial \phi}+\frac{\partial^2 g_{\phi}^2\varrho}{\partial \phi^2}\right)\,. \label{eq:fpe-sm}
\end{align}
\end{widetext}
We are interested in the evolution of $A$ (and $h$, $g_A$, $g_{\phi}$, and $\varrho$ are all $2\pi$-periodic in $\phi$), so we integrate both sides of the FPE (\ref{eq:fpe-sm}) from $\phi = 0$ to $\phi = 2\pi$ to obtain
\begin{align}
	\frac{\partial Q}{\partial t} &= -\frac{\partial}{\partial A}\left[h(A) Q\right] +\frac{\sigma^2}{2} \frac{\partial^2}{\partial A^2}\left(\int_0^{2\pi} g_A^2 q d\phi\right)\notag \\ &-\frac{\sigma^2}{2}\frac{\partial}{\partial A}\int_0^{2\pi}\left(g_A\frac{\partial g_A}{\partial A}+g_{\phi}\frac{\partial g_{\phi}}{\partial \phi}\right)q d\phi \,, \label{eq:fpe-sm2}
\end{align}
where $Q(A,t) = \int_0^{2\pi}\varrho(A,\phi,t)d\phi$ is the PDF of $A$ averaged over $\phi$.  Note that the integral in (\ref{eq:fpe-sm2}) amounts to averaging over a ``fast" variable.



We then perform averaging based on the assumption \cite{nakaoetal,nagaikori} that $Q(A,t)$ evolves slowly compared to the time scale of oscillations. The resulting FPE has a steady state given by
\begin{equation}
	Q_{\infty}(A)=\frac{C}{P_{1}(A)}\exp\left(\int\frac{[2h(A)+\sigma^2P_2(A)]dA}{\sigma^2P_{1}(A)}\right)\,, \label{eq:11}
\end{equation}
where 
\begin{align}
	h(A) &= (K-2\gamma)A-KA^2\,, \notag \\
	P_1(A) &:= \frac{1}{T}\int_{0}^{T}g_A^2 dt = \frac{1}{2}(1-A)^2\sum_{n=1}^{\infty}A^n (|c_n|^2)\,, \notag \\
	P_2(A) &:= \frac{1}{T}\int_{0}^{T}g_A \frac{\partial g_A }{\partial A} dt + \frac{1}{T}\int_{0}^{T}g_\phi \frac{\partial g_A }{\partial \phi} dt \notag \\ &= \frac{1}{2}(1-A)\sum_{n=1}^{\infty}A^{n-1}(n-A)(|c_n|^2) \,, 
\end{align}	
and $C$ is a constant obtained from the normalization $\int_0^1 Q_{\infty}=1$. 


\section{Generalized Cauchy Distribution Of Frequencies} 

We apply the above results to extend the theory developed in Ref.~\cite{nagaikori} to generalized Cauchy distributions of oscillator frequencies. We set $\cZ=\sin(\theta)$, so $b_1=1$ and all other Fourier coefficients vanish. This yields $P_{1}(A)=A(1-A)^2/2$ and $P_2(A)=(1-A)^2/2$ \footnote{Therefore, our results are identical for any function $\cZ$ whose only nonvanishing Fourier coefficients satisfy $|c_1|^2 = 1$.}. The order parameter signifying the transition between synchrony and asynchrony adopted in Ref.~\cite{nagaikori} is the maximum of the PDF $Q_{\infty}$. 
To find where $Q_{\infty}$ attains its maximum, we set $Q'_\infty=0$. This yields
\begin{equation}
	\frac{2h(A)}{\sigma^2P_{1}(A)^2}+\frac{P_2(A)}{P_1(A)^2}-\frac{P'_{1}(A)}{P_{1}(A)^2}=0\,. \label{eq:12}
\end{equation}
Using our expressions for $h(A)$, $P_1(A)$, and $P_{2}(A)$ then gives $A=\min\left\{0,\frac{K+\frac{\sigma^2}{2}-2\gamma}{K+\frac{\sigma^2}{2}}\right\}$, so we need $K+\sigma^2/2>2\gamma$ for
 synchrony.



\begin{figure}[h!]
\includegraphics[width=0.23\textwidth]{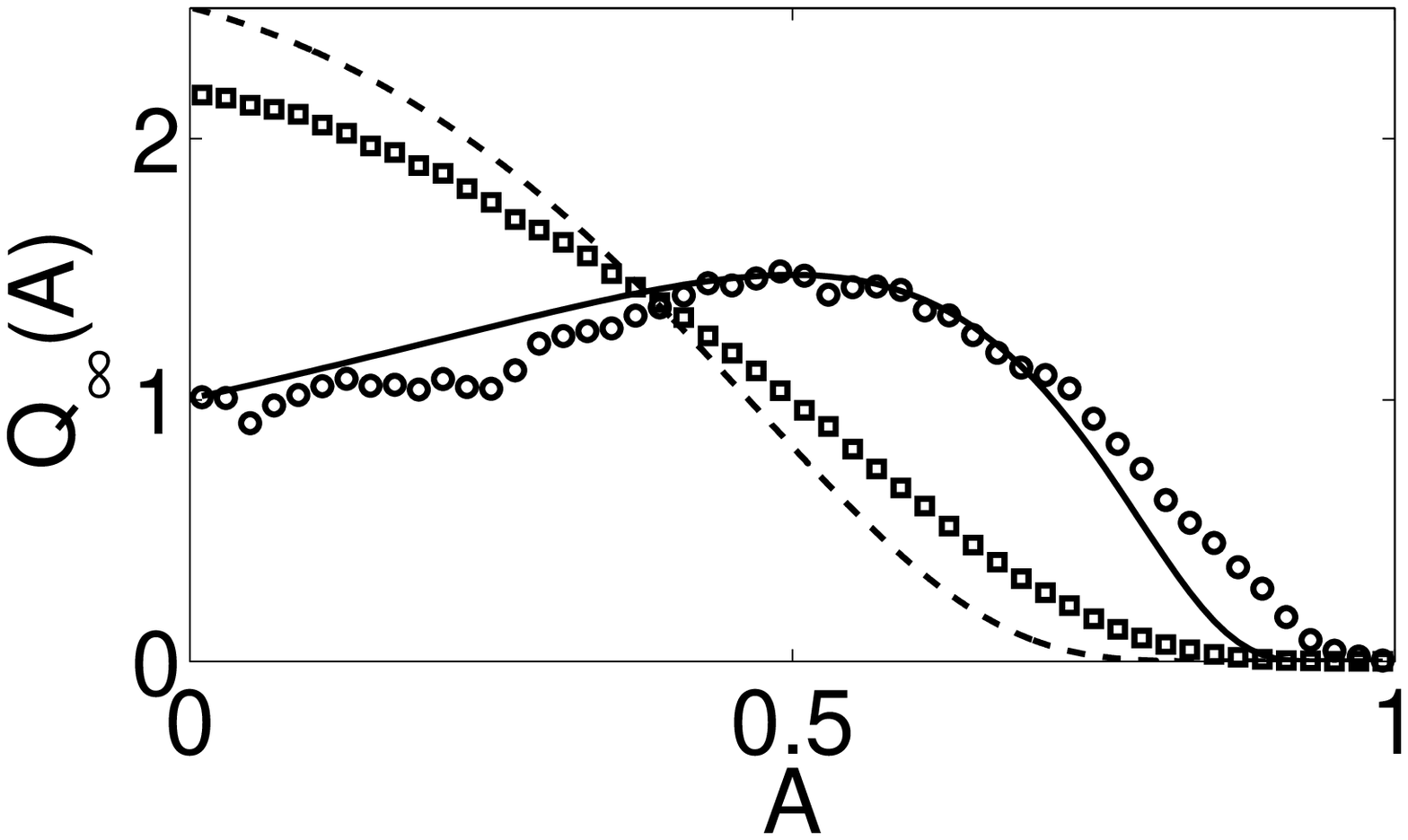}
\includegraphics[width=0.23\textwidth]{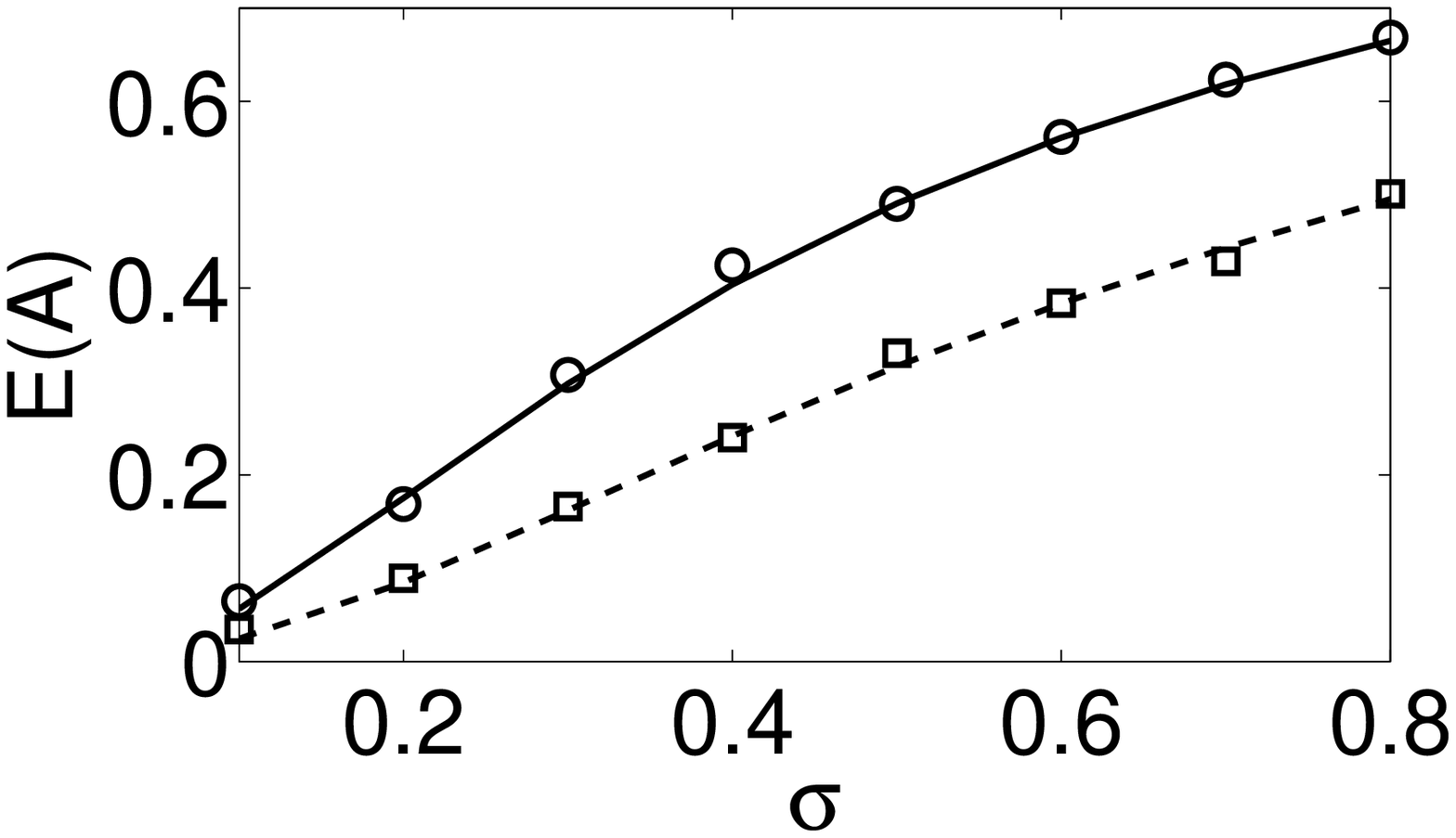}
\caption{(Left) Plots of the PDF $Q_{\infty}(A)$. We calculate curves from the analytical expression (\ref{eq:12}), and we plot circles and squares are from 50-bin histograms of data obtained from one realization of direct numerical simulations. The solid curve and circles are for the case $\sigma=0.4$ and $\gamma=0.02$, and the dashed curve and squares are for $\sigma=0.4$ and $\gamma=0.05$. (Right) Plots of the measure of synchrony $\E(A)$ versus $\sigma$. We obtain the curves from analytical calculations (\ref{eq:12}), and the circles and squares represent data from 
a temporal average of one realization. The solid curve and circles are for $\gamma=0.02$, and the dashed curve and squares are for $\gamma=0.05$.
}
\label{fig:1}
\end{figure}

The aforementioned techniques can be applied to many scenarios. The case in which $\sigma=0$ has been studied \cite{cite7}, and Ref.~\cite{nagaikori} provides a detailed discussion for $\gamma=1$.  Let's consider the case $K = 0$ in which uncoupled, nonidentical oscillators are driven by noise. Several studies have considered a noise-driven ensemble of identical oscillators \cite{nakaoetal,bresslofflai11,cite5}, but there has been much less work on nonidentical oscillators. 
We begin with the case $\cZ(\theta)=\sin(\theta)$ to simplify our expression for $Q_\infty$ in equation (\ref{eq:11}) to obtain 
\begin{equation}
	Q_{\infty}(A)=\frac{2C}{(1-A)^2}\exp\left(-\frac{8\gamma}{\sigma^2(1-A)}\right)\,. 
\end{equation}	
We expect to observe a peak at $A>0$ for $\sigma^2-4\gamma>0$ and a peak at $A=0$ for $\sigma^2-4\gamma \leq 0$.
We confirm this prediction by simulating an ensemble of $N=10000$ phase oscillators evolving according to equation (\ref{eq:1}). We constructed the generalized Cauchy distribution for the natural frequencies using 
\begin{equation*}
	\omega_j=\omega_0+\gamma \tan{\frac{\pi}{2N}[2j-(N+1)]} 
\end{equation*}	
for the $j$th 
oscillator \cite{daido86}.  In Fig.~\ref{fig:1}, we compare the computed PDF $Q_{\infty}(t)$ with histograms of $A$ that we obtained from direct numerical simulations (i.e., using stochastic simulations).  Observe that we obtain a peak at $A>0$ for $\sigma^2-4\gamma \approx 0.08$ but a peak at $A = 0$ for $\sigma^2-4\gamma \approx -0.04$.  We obtain good qualitative agreement between $Q_{\infty}(t)$ and $A$, though the noisy nature of the system entails some mismatch between theory and direct simulations. 

The increase in synchrony is gradual 
as $\sigma^2-4\gamma$ changes signs. Accordingly, in addition to using the position of the peak to measure synchrony, we also use $\E(A)=\int_0^1 AQ_{\infty}(A)dA$.  
We show our results in the right panel of Fig.~\ref{fig:1}. Using both theory and simulations, we see that $\E(A)$ increases with the strength of the common noise and decreases with the width of the distribution.  As Fig.~\ref{fig:1} illustrates, even systems with only $N = 50$ oscillators already exhibit very good agreement for the expectation $\E(A)$.

\section{General Phase-Sensitivity Functions} 

We wish to study the effects of noise via a general phase-sensitivity function $\cZ(\theta)$ rather than just $\cZ(\theta) = \sin(\theta)$. 
This is relevant for phase oscillator models arising from dynamical systems in fields like physics and biology \cite{scholarsync}. 
A sinusoidal phase-sensitivity function is overly simplistic \cite{scholarsync}, but one can approximate many functions $\cZ(\theta)$ using only a few terms in its Fourier series. Nakao et al. \cite{nakaoetal} showed for uncoupled, identical limit-cycle oscillators that higher harmonics of $\cZ$ can cause oscillator ensembles to form clusters around a limit cycle and that increasing the strength of common noise makes the oscillators more sharply clustered (i.e., their phases reside in a smaller interval).
Equally-spaced (or almost equally-spaced) clusters lead to cancellation effects and a decrease in the value of the order parameter $|r(t)|$, which is problematic for our previous analysis. Moreover, the formation of multiple clusters causes the OA ansatz to break down: from the normalization of the phase distribution $f(\omega,\theta,t)$, we know that $|\alpha|<1$, so the coefficients of higher modes must have smaller magnitude than that of the first mode.  Thus, we do not get a phase distribution with multiple clusters.
(For example, to obtain three equally spaced clusters, one would expect the third mode to have the largest-magnitude coefficient.)



\begin{figure}[h!]
\includegraphics[width=.23\textwidth]{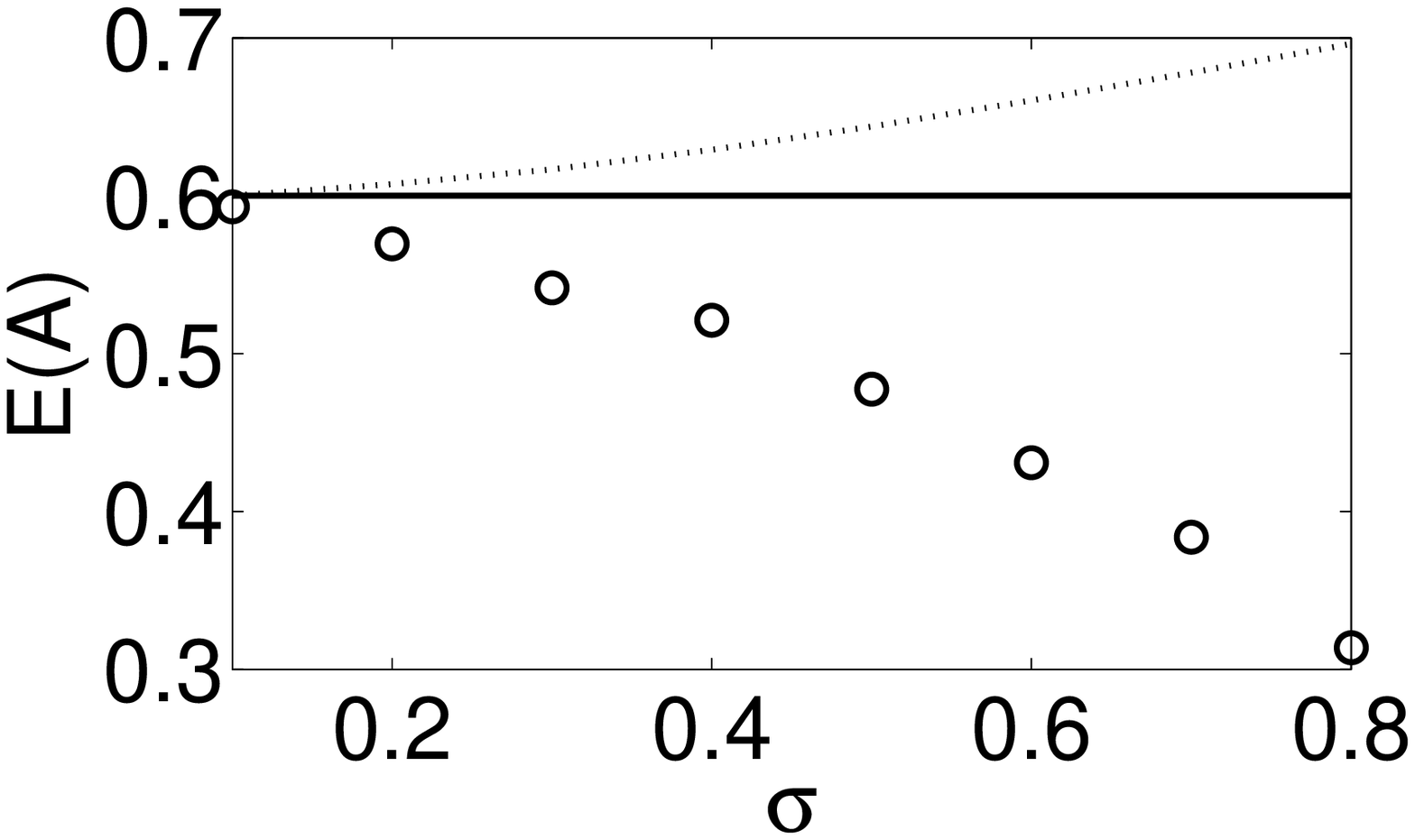}
\includegraphics[width=.23\textwidth]{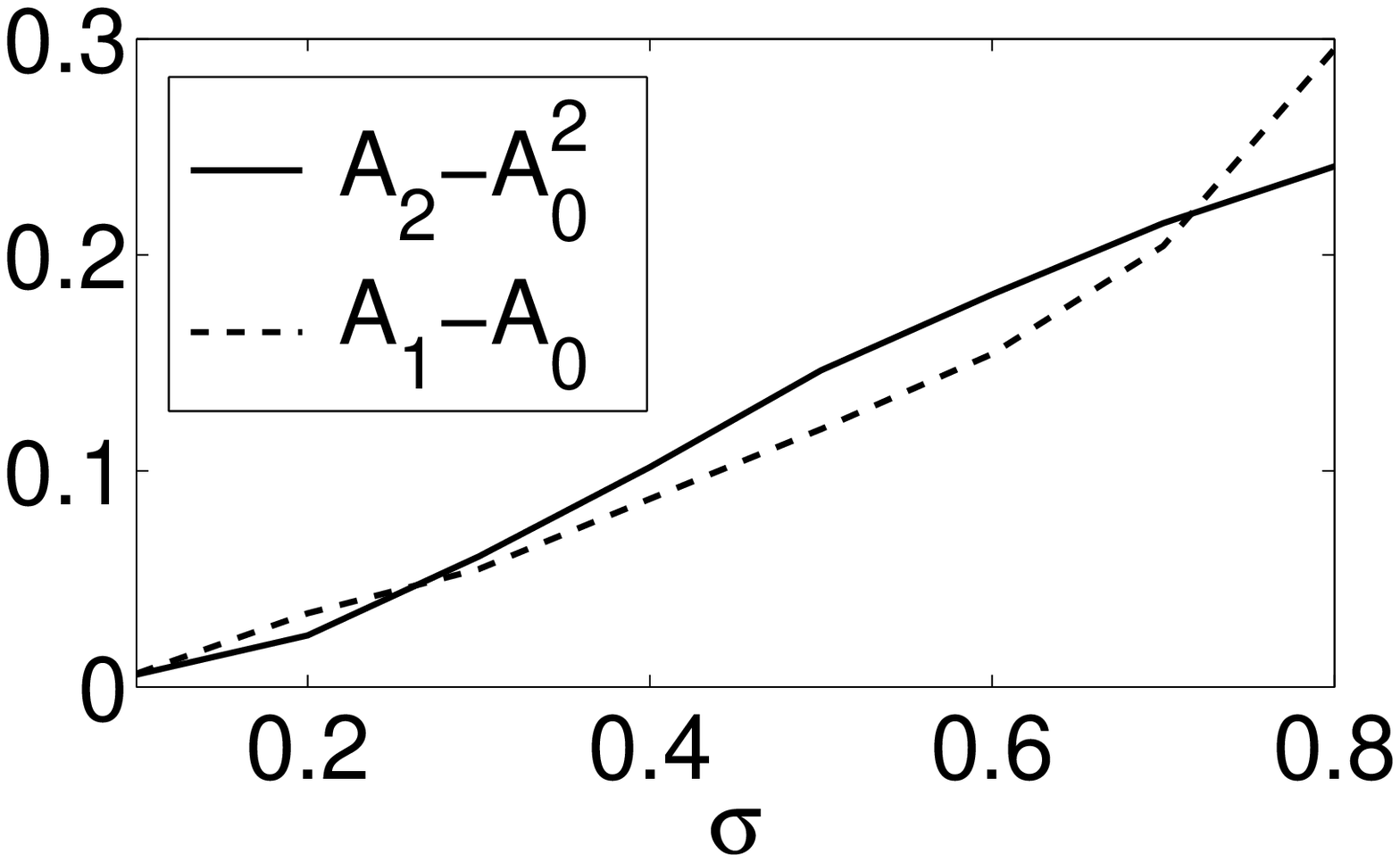}
\caption{(Left) The synchrony measure $\E(A)$ versus $\sigma$ for $K=0.5$ and $\gamma=0.1$. The dashed curve is our erroneous calculation of synchrony, the solid curve is our estimate of synchrony in the absence of noise, and the circles are from direct numerical simulations. (Right) Comparison of the clustering effect $\langle A_2\rangle-A_0^2$ and the noise-induced decrease in synchrony $A_0-\langle A_1\rangle $ from the left panel.
}
\label{fig:3}
\end{figure}

To illustrate the breakdown of the OA ansatz, we consider the example $\cZ(\theta)=\sin(2\theta)+\cos(2\theta)$. This function can arise in an ensemble of Stuart-Landau oscillators from adding multiplicative common noise (where the noise strength is multiplied by a function of one or more system variables), as in one of the case studies in Ref.~\cite{nakaoetal}.
 This yields $P_1=A^2(1-A)^2/2$ and $P_2=A(1-A)(2-A)/2$, which we insert into equation (\ref{eq:11}) to calculate the steady-state pdf $Q_\infty{(A)}$ and the order parameter $\E(A)$. We also estimate the level of synchrony in the absence of noise by setting $\sigma=0$.  (We use the notation $A_0$ to denote values of $A$ in this situation.) This yields $h(A_0)Q=\mathrm{constant}$.  Consequently, $Q(A_0)$ diverges at the zeros of $h(A_0)$, which occur at $A_0 = 0$ and $A_0=1-2\gamma/K$. We show our numerical results in the left panel of Fig.~\ref{fig:3}. Observe that the presence of the higher harmonic leads to a decrease in synchrony rather than an increase in synchrony with increased noise strength, in contrast to many studies of noise-induced synchrony \cite{cite6,cite6b,nagaikori}. 

To characterize this decrease in synchrony, we use a family of order parameters from Ref.~\cite{skardaletal} to study clustering.  We define $A_2=|r_2|^2$, where 
\begin{equation}
	r_2(t)=\int_{-\infty}^{\infty}\int_0^{2\pi} \exp(2i\theta) f(\omega,\theta,t) d\theta d\omega\,.  
\end{equation}	
One can similarly define $A_q$ for all $q \in \mathbb{Z^+}$.
For the OA ansatz to hold, one needs $A_2=A_0^2$. As we show in the right panel of Fig.~\ref{fig:3} using direct numerical simulations, we find a high correlation between the clustering effect quantified by $\langle A_2\rangle-A_0^2$ and the noise-induced decrease in synchrony quantified by $A_0-\langle A_1\rangle$.  (The notation $\langle x \rangle$ refers to the temporal average of the variable $x$.)

\section{Clustering} 

\begin{figure}[h!]
\includegraphics[width=.23\textwidth]{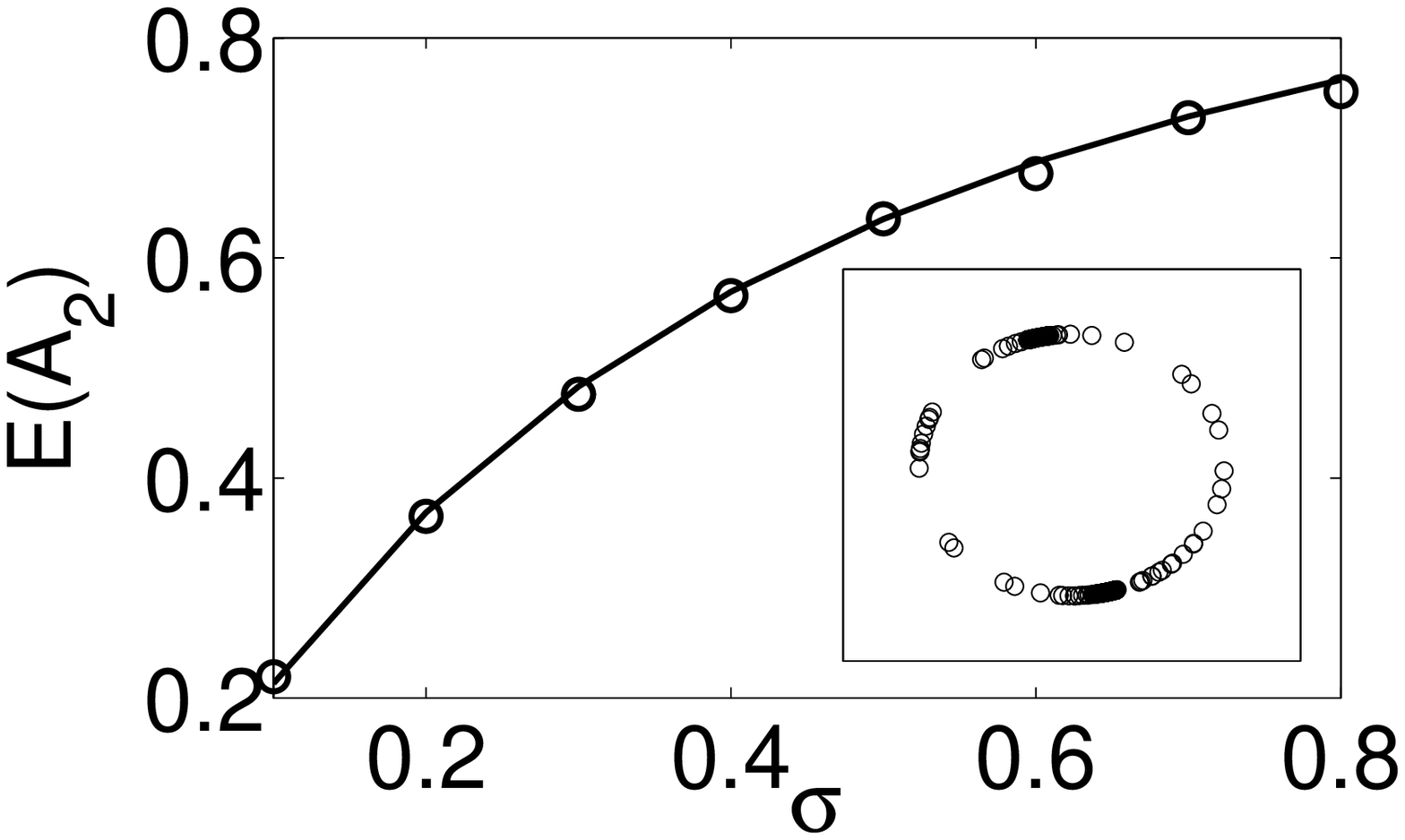}
\includegraphics[width=.23\textwidth]{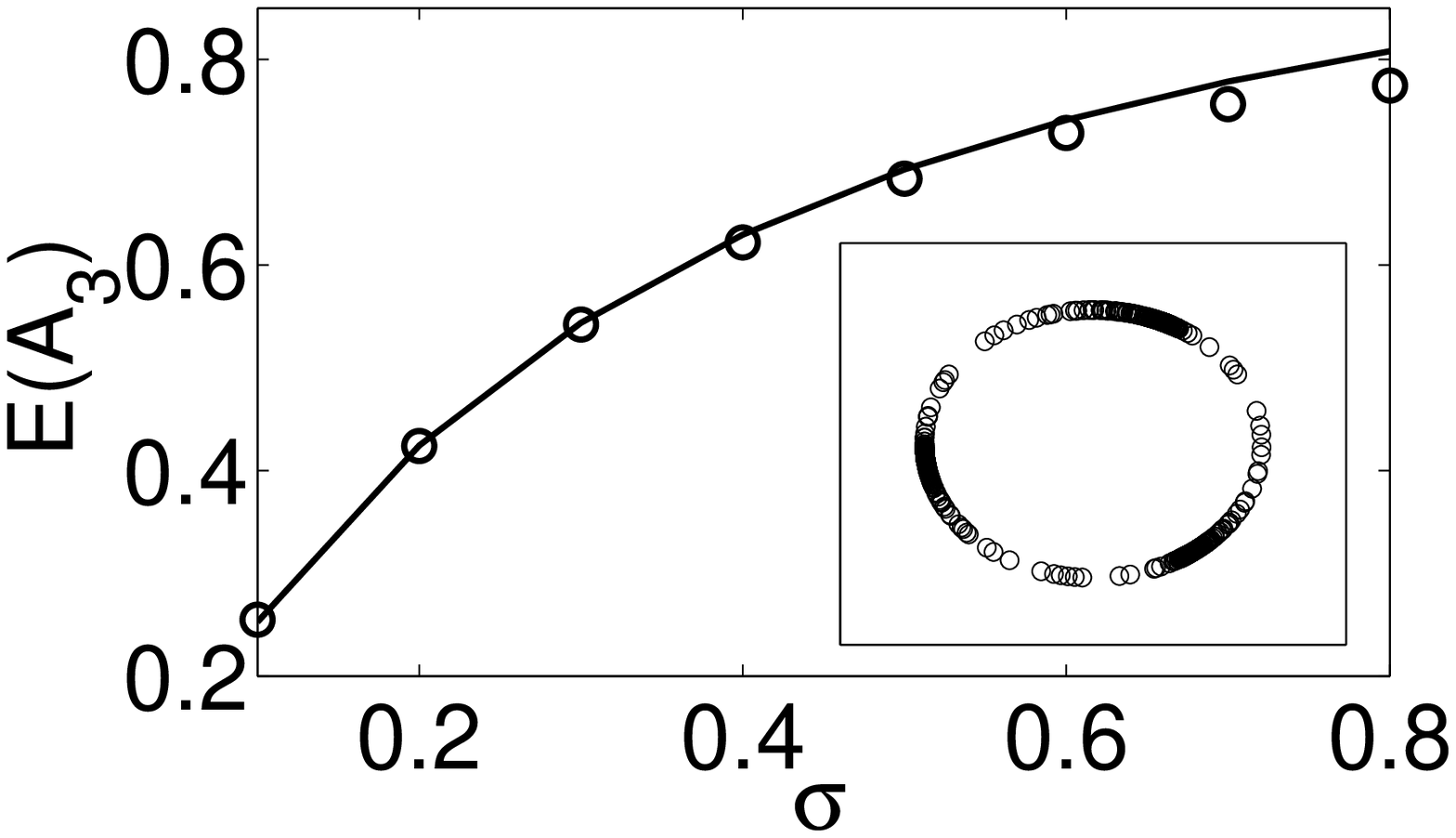}
\caption{Cluster synchrony induced by noise and coupling via the phase-sensitivity function $\cZ=\sin(q\theta)+\cos(q\theta)$ and coupling between oscillators of the form $\sin(q[\theta_j-\theta_i])$.  We use the parameter values $K=0.1$ and $\gamma=0.05$. The solid curves are from analytical calculations, and the circles are from direct numerical simulations. The insets show snapshots of $N = 500$ oscillators for $\sigma=0.8$. The left panel is for $q=2$, and the right panel is for $q=3$. 
}
\label{fig:5}
\end{figure}

We now show that noise increases cluster synchrony when there is higher-order coupling (i.e., when the dominant mode in the coupling function is not the 
$q = 1$ Fourier mode).
We take 
\begin{align*}
	\cZ_q &= a_q\sin(q\theta)+b_q\cos(q\theta) \\
		&= c_q\exp\left\{qi\theta\right\}+c^*_q\exp\left\{-qi\theta\right\}\,, \quad q \in \mathbb{Z^+}
\end{align*}	 
to obtain
\begin{align}
	\frac{d\theta_i}{dt}=\omega_i+\frac{K}{N}\sum_{j=1}^N \sin(q[\theta_j-\theta_i])+\sigma\cZ_q(\theta_i)p(t)\,,
\end{align}
which was discussed for the case $\sigma=0$ in Ref.~\cite{skardaletal}. By defining the mode-$q$ order parameter 
\begin{equation}
	r_q(t)=\int_{-\infty}^{\infty}\int_0^{2\pi} \exp(qi\theta) f(\omega,\theta,t) d\theta d\omega\,, 
\end{equation}
we derive
\begin{equation}
	\frac{\partial f}{\partial t}+\frac{\partial}{\partial \theta}\left\{\left[\omega+Kr_q\sin(q\theta)+\sigma \cZ_q p(t)\right]f\right\}=0\,,  
\end{equation}	
which is similar to the nonlinear PDE (\ref{eq:2}) that we obtained above. Applying the same method as before yields
\begin{equation}
	\frac{dr_q}{dt}=q\left[(-\gamma+i\omega_0) r_q+\frac{K}{2}r_q(1-|r_q|^2)+i\sigma p(c_q^*+r_q^2 c_q)\right]\,.
\end{equation}
Setting $r_q=A_q\exp(qi\theta)$ and following the previously discussed procedure yields the steady-state PDF 
\begin{equation}
	Q_{\infty}(A_q)=\frac{C}{P_{1}(A_q)}\exp\left(\int\frac{[2h(A_q)+q\sigma^2P_2(A_q)]dA_q}{q\sigma^2P_{1}(A_q)}\right)\,, 
\end{equation}	
where $P_{1}(A_q)=A_q(1-A_q)^2(|c_q|^2)/2$ and $P_2(A_q)=(1-A_q)^2(|c_q|^2)/2$.  This, in turn, implies that noise and coupling both increase the ``$q$-cluster synchrony" of the system. We verify this for two and three clusters in Fig.~\ref{fig:5}.




\section{Antiferromagnetic Coupling} 

We now consider interactions of noise and coupling for oscillator systems with \emph{antiferromagnetic coupling}, in which there are two groups of oscillators with positive coupling between oscillators in the same group but negative coupling between oscillators in different groups. We label the two groups as ``odd" and ``even" oscillators. The temporal evolution of the phase of the $i$th oscillator is 
\begin{equation}
	\frac{d\theta_i}{dt}=\omega_i+\frac{1}{N}\sum_{j=1}^N K_{ij}\sin(\theta_j-\theta_i)+\sigma\cZ(\theta_i)p(t)\,, \label{eq:af1}
\end{equation}
where $K_{ij}=K$ if $i+j$ is even and $K_{ij} = -K$ if it is odd. We show that the oscillators form two distinct clusters when $K > K_c = 2\gamma$ in the absence of noise (i.e., for $\sigma=0$).  We define an \emph{antiferromagnetic order parameter} 
\begin{equation}
	r_{\mathrm{af}}(t)=(1/N)\sum_j(-1)^j\exp(i\theta_j) 
\end{equation}	
and demonstrate that the dependence of $|r_{\mathrm{af}}|$ on $K$ and $\gamma$ is analogous to what occurs in the conventional Kuramoto model. 

By considering odd oscillators and even oscillators as separate groups of oscillators, we define the complex order parameters
\begin{align}
	r_{o} &= \frac{2}{N}\sum_j^{N/2} \exp(i \theta_{2j-1})\,, \notag \\
	r_{e} &= \frac{2}{N}\sum_j^{N/2} \exp(i \theta_{2j})
\end{align}
for the odd and even oscillators, respectively (also see Ref.~\cite{laing12}). The antiferromagnetic order parameter can then be expressed as $r_{\mathrm{af}}=(r_o+r_e)/2$. As with the usual global, equally weighted, sinusoidal coupling in the Kuramoto model (which we call \emph{ferromagnetic coupling}), we let the number of oscillators $N \rightarrow \infty$ and examine continuum oscillator densities $f_{o,e}(\omega,\theta,t)$. Following the analysis for the Kuramoto model in Ref.~\cite{cite7}, the continuity equations for the densities of the oscillators take the form of a pair of nonlinear FPEs:
\begin{equation}
	\frac{\partial f_{o,e}}{\partial t}+\frac{\partial}{\partial \theta}\left[\left(\omega+\frac{K}{2}r_{o,e}\sin(\theta)-\frac{K}{2}r_{e,o}\sin(\theta)\right)f_{o,e}\right]=0\,.
\end{equation}

One can then apply Kuramoto's original analysis \cite{kuramoto84} to this system. Alternatively, one can proceed as in the ferromagnetic case and apply the OA ansatz separately to each family of oscillators. One thereby obtains the coupled ordinary differential equations (ODEs)
\begin{align}\label{nineteen}
	\frac{dr_{o,e}}{dt} =&-\gamma r_{o,e}+ i\omega_0 r_{o,e}\notag \\&+\frac{K}{4}\left[(r_{o,e}-r_{e,o})-r_{o,e}^2(r_{o,e}^*-r_{e,o}^*)\right]\,.
\end{align}
Taking the sum and difference of the two equations in (\ref{nineteen}) yields
\begin{widetext}
\begin{align}
	\frac{d(r_e-r_o)}{dt} &= \left(-\gamma+i\omega_0+\frac{K}{2}\right)(r_e-r_o)+\frac{K}{4}\left(-r_e^2r_e^*+r_e^2r_o^*+r_o^2r_o^*-r_o^2r_e^*\right)\,, \notag \\
	\frac{d(r_e+r_o)}{dt} &= -\gamma(r_e+r_o)+\frac{K}{4}\left(-r_e^2r_e^*+r_e^2r_o^*-r_o^2r_o^*+r_o^2r_e^*\right)\,. \label{eq:14}
\end{align}
\end{widetext}
In the case of ferromagnetic coupling, we let $r=\sqrt{A}\exp(\omega_0 t+\phi)$. If one were to proceed analogously in antiferromagnetic coupling and define $r_{o,e}=\sqrt{A_{o,e}}\exp(\omega_0 t+ \phi_{o,e})$, one would obtain four coupled SDEs for $A_{o,e}$ and $\phi_{o,e}$, and it is then difficult to make analytical progress. However, we seek to quantify the aggregate level of synchrony only in the absence of noise.  In this case, after initial transients, 
steady states satisfy
$A_e=A_o$ and $\phi_e=-\phi_o=\psi/2$, where $\psi$ is the phase difference between the two groups. (We cannot use this method in the presence of noise, as noise breaks the symmetry.)

Equations (\ref{eq:14}) then simplify to
\begin{widetext}
\begin{align}
	\frac{dA}{dt}\sin\left(\frac{\psi}{2}\right)+A\cos\left(\frac{\psi}{2}\right)\frac{d\psi}{dt}&=-2\gamma A\sin \left(\frac{\psi}{2}\right)+KA\sin\left(\frac{\psi}{2}\right)+\frac{1}{2}KA^2\left[\sin\left(\frac{3\psi}{2}\right)-\sin\left(\frac{\psi}{2}\right)\right]\,, \notag\\
	\frac{dA}{dt}\cos\left(\frac{\psi}{2}\right)-A\sin\left(\frac{\psi}{2}\right)\frac{d\psi}{dt}&=-2\gamma A\cos\left(\frac{\psi}{2}\right)+\frac{1}{2}KA^2\left[\cos\left(\frac{3\psi}{2}\right)-\cos\left(\frac{\psi}{2}\right)\right]\,.
\end{align}
\end{widetext}
This, in turn, yields
\begin{align}
	\frac{dA}{dt}&=-2\gamma A+KA(1-A)\sin^2\left(\frac{\psi}{2}\right)\,, \label{eq:15} \\
	\frac{d\psi}{dt}&=\frac{1}{2}K(1+A)\sin\psi\,. \label{eq:16}
\end{align}

\begin{figure}[h!]
\includegraphics[width=.49\textwidth]{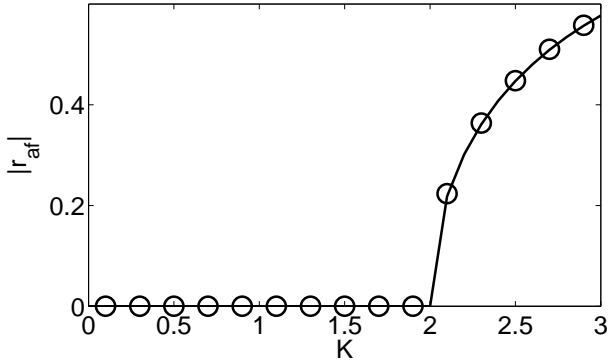}
\caption{Antiferromagnetic order parameter $|r_{\mathrm{af}}|$ versus coupled strength $K$ for width parameter $\gamma=1$ in the absence of noise (i.e., for $\sigma=0)$.
 The solid curve is the analytical steady state, and circles are from direct numerical simulations of the ODEs for an ensemble of $N = 500$ oscillators.
}
\label{fig:7}
\end{figure}

By setting $\frac{dA}{dt}=\frac{d\psi}{dt}=0$, we seek equilibria of the system. When $K\sin^2(\psi/2)\leq2\gamma$, there is an unstable equilibrium at $(A,\psi) = (0,0)$ and a stable equilibrium at $(A,\psi) = (0,\pi)$. When $K\sin^2(\psi/2)>2\gamma$, this equilibrium point is unstable. Additionally, there is an unstable equilibrium at $(A,\psi) = \left(1-\frac{2\gamma}{K\sin^2(\psi/2)},0\right)$ and a stable equilibrium at $(A,\psi) =\left (1-\frac{2\gamma}{K\sin^2(\psi/2)},\pi\right)$. In practice, this implies that $\psi(t) \rightarrow \pi$, so the threshold for observing synchrony is $K_c=2\gamma$ (just as in the Kuramoto model). Similarly, the antiferromagnetic order parameter $|r_{\mathrm{af}}|=\sqrt{A}\sin\left(\psi/2\right)$ has a stable steady state at $\min\{0,\sqrt{1-K_c/K}\}$, which has the same dependence on $K$ as the Kuramoto order parameter does in the traditional Kuramoto model \cite{kuramoto84,cite7}. We plot the antiferromagnetic order parameter versus the coupling strength $K$ in Fig.~\ref{fig:7} and obtain excellent agreement with direct numerical simulations of the coupled oscillator system.

\begin{figure}[h!]
\includegraphics[width=.23\textwidth]{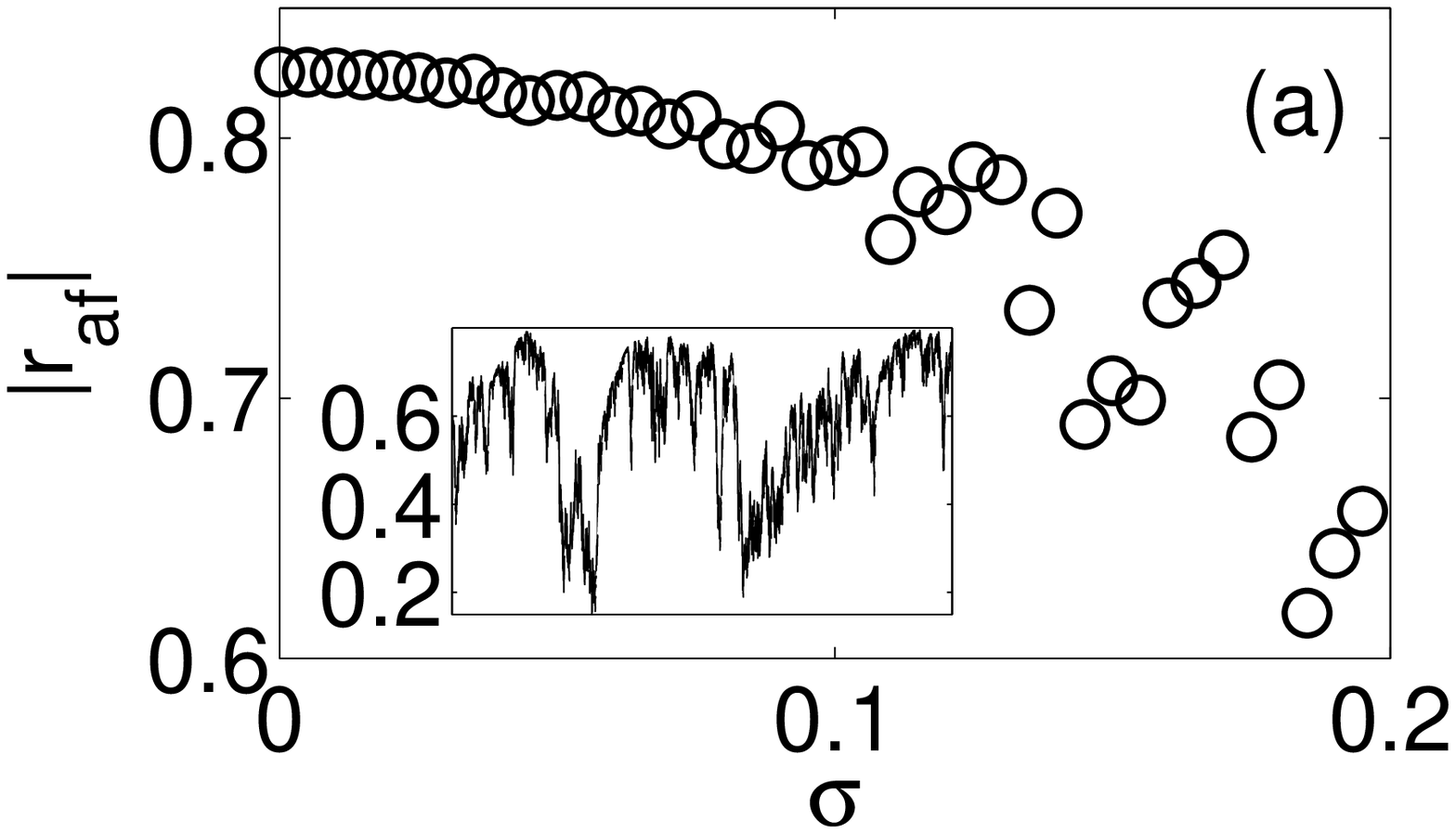}
\includegraphics[width=.23\textwidth]{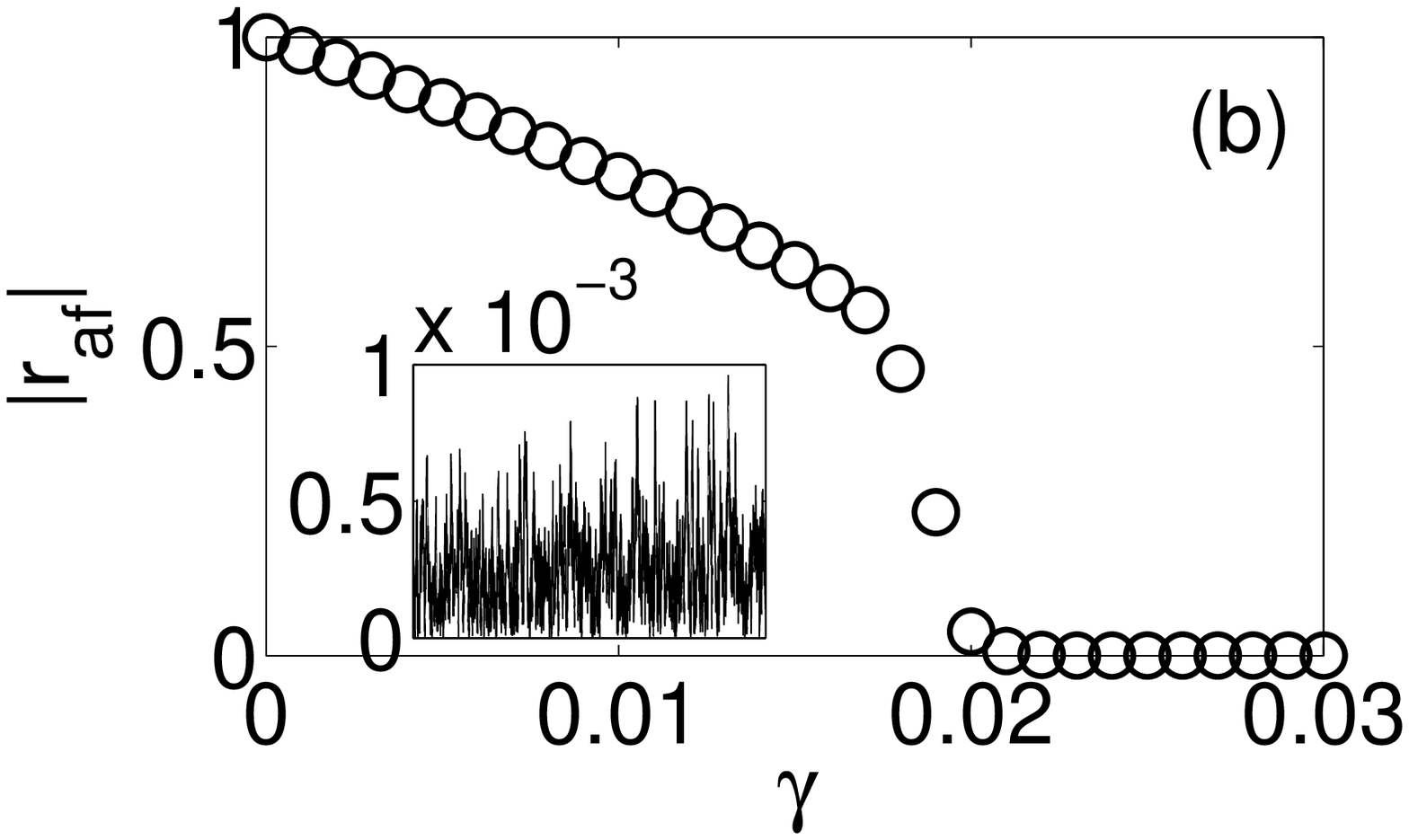}
\includegraphics[width=.23\textwidth]{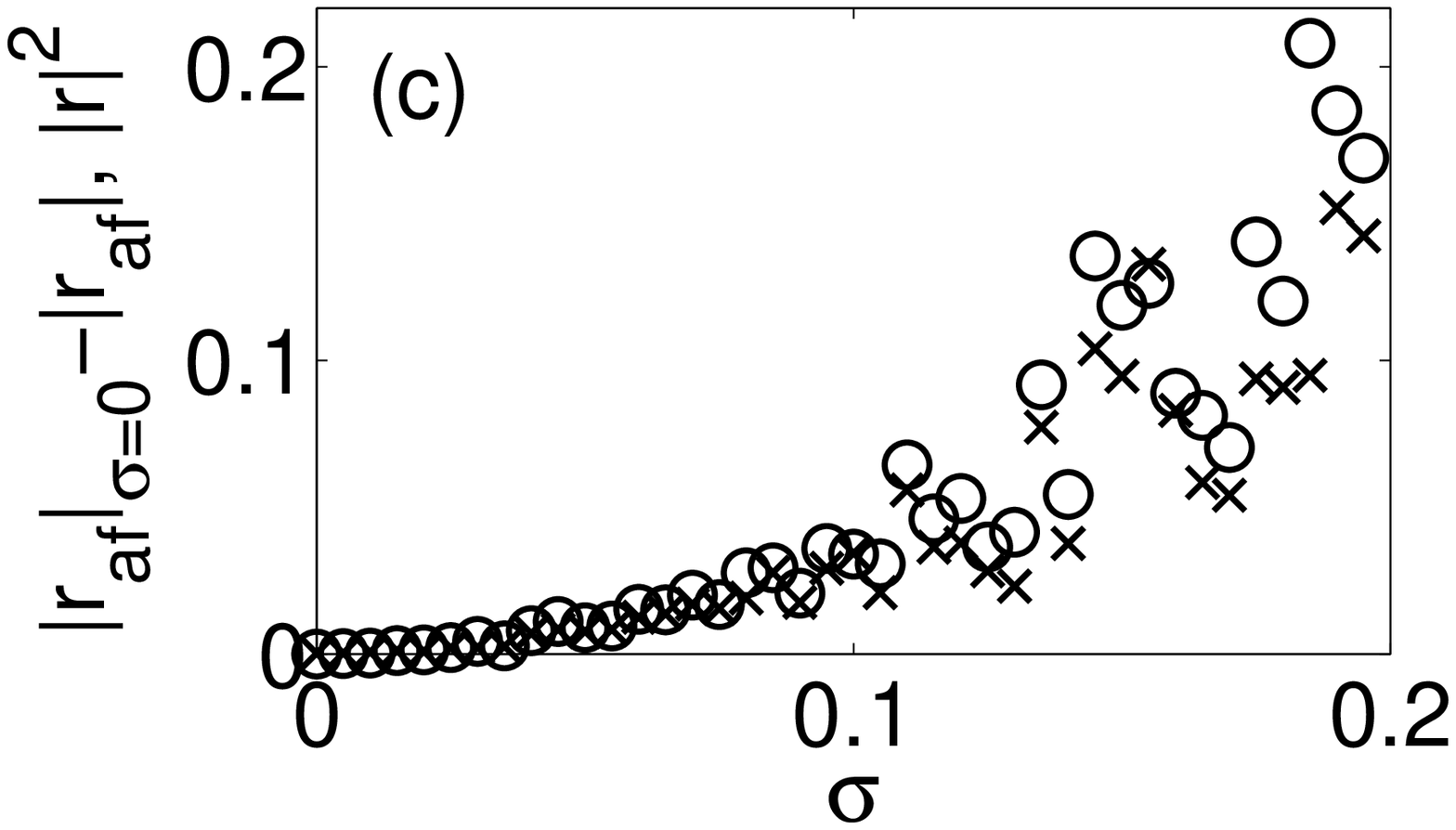}
\includegraphics[width=.23\textwidth]{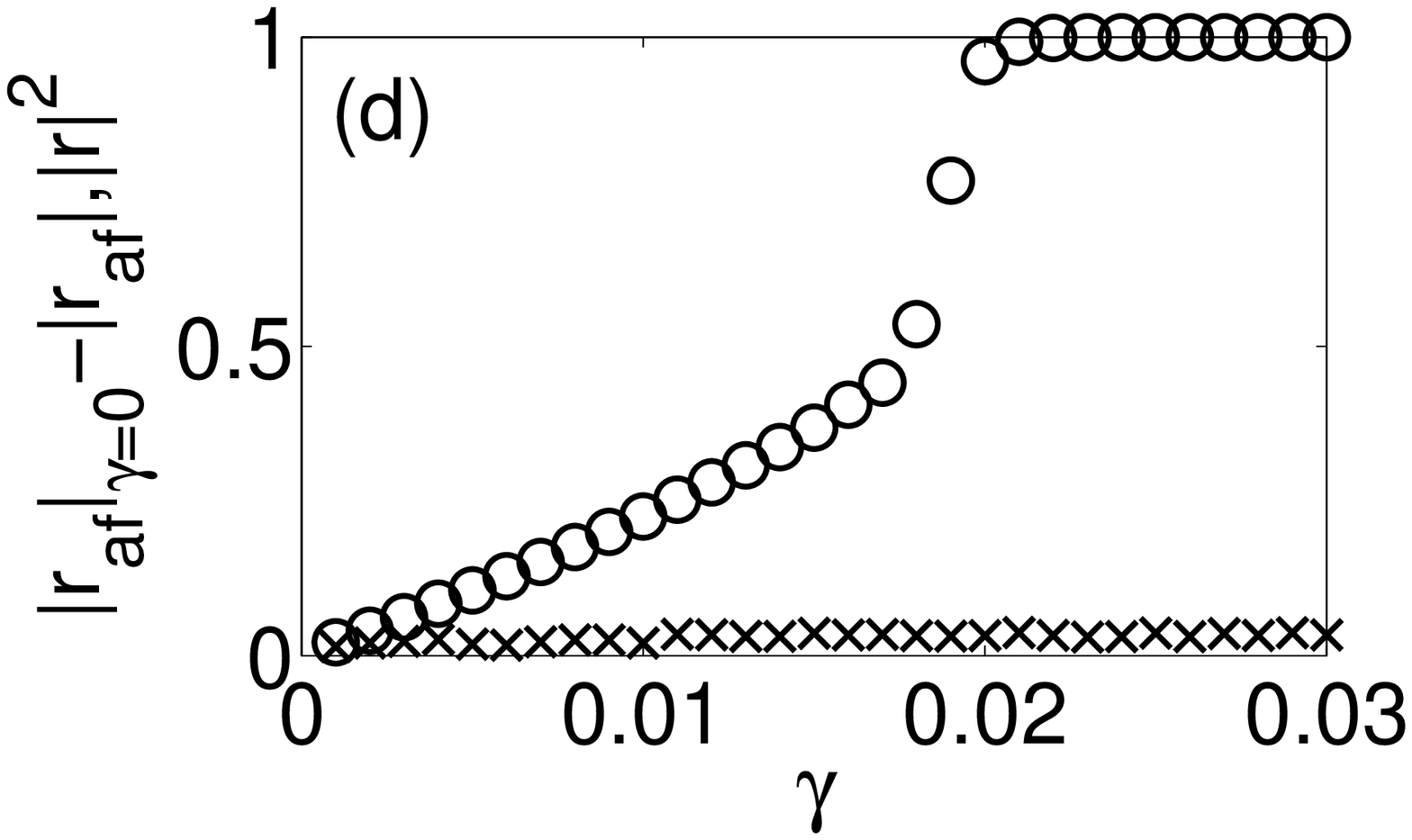}
\caption{Results of direct numerical simulations for antiferromagnetically coupled phase oscillators. (a) Antiferromagnetic synchrony $|r_{\mathrm{af}}|$ versus noise strength $\sigma$ for $K=0.05$ and $\gamma=0.008$. In the inset, we show a sample realization for $\sigma=0.5$ between times $t=1000$ and $t=2000$. (b) Antiferromagnetic synchrony $|r_{\mathrm{af}}|$ versus $\gamma$ for $K=0.05$ and $\sigma=0.01$. In the inset, we show a sample realization for $\gamma=0.05$ between times $t=1000$ and $t=2000$. (c) Circles give the decrease of antiferromagnetic synchrony ($|r_{\mathrm{af}}|_{\sigma=0}-|r_{\mathrm{af}}|$), and crosses give the square of the usual Kuramoto measure of synchrony $|r|^2$. (d) Same as panel (c), except the horizontal axis is the natural frequency distribution width parameter $\gamma$ rather than $\sigma$. [Each data point in the figures in the main panels represents the temporal average of (\ref{eq:af1}) with $N=500$ oscillators.]
}
\label{fig:6}
\end{figure}

We now consider the effect of correlated noise on the system (\ref{eq:af1}). As we have seen previously, the effect of noise when the first Fourier mode of $\cZ$ dominates is to synchronize the oscillators (i.e., to form a single cluster). 
In Fig.~\ref{fig:6}, we explore this using direct numerical simulations.



In agreement with our intuition, the noise and coupling have contrasting effects. Accordingly, the antiferromagnetic synchrony $|r_{\mathrm{af}}|$ decreases with increasing noise strength $\sigma$ (see Fig.~\ref{fig:6}a). As shown in the inset, the noise causes the system to ``jump" between states with low and high values of $|r_{\mathrm{af}}|$.  By contrast, as shown in Fig.~\ref{fig:6}b, $|r_{\mathrm{af}}|$ decreases with increasing natural frequency distribution width parameter $\gamma$. Additionally, the decrease in synchrony, $|r_{\mathrm{af}}|_{\sigma=0}-|r_{\mathrm{af}}|$, correlates positively with the increase in the traditional measure of synchrony $|r|=|(1/N)\sum_j\exp(i\theta_j)|=\sqrt{A}$ (see Fig.~\ref{fig:6}c). (The Pearson correlation coefficient between $|r_{\mathrm{af}}|_{\sigma=0}-|r_{af}|$  and $|r|^2$ is $0.955$.) There is no such  relationship in the case in which $\gamma$ is increased, as $|r|$ remains small and approximately constant (see Fig.~\ref{fig:6}d).


\section{Conclusion} 

We have examined noise-induced synchronization, desynchronization, and clustering in globally coupled, nonidentical oscillators. We demonstrated that noise alone is sufficient to synchronize nonidentical oscillators.
 However, the extent to which common noise induces synchronization depends on the magnitude of the coefficient of the first Fourier mode. In particular, the domination of higher Fourier modes can disrupt synchrony by causing clustering.  
We then considered higher-order coupling and showed that the cluster synchrony generated by such coupling is reinforced by noise if the phase-sensitivity function consists of Fourier modes of the same order as the coupling.  




One obvious avenue for future work is the development of a theoretical framework that would make it possible to consider multiple harmonics of both the coupling function and the phase-sensitivity function. It would also be interesting to consider generalizations of antiferromagnetic coupling, such as the variant studied in Ref.~\cite{laing12}. One could also examine the case of uncorrelated noise, 
which has been studied extensively \cite{cite8} via an FPE of the form 
\begin{equation*}
	\frac{\partial f}{\partial t}+\frac{\partial}{\partial \theta}\left[\omega+K r\sin(\theta)\right]=\frac{\partial^2 f}{\partial \theta^2}\,. 
\end{equation*}	
However, proceeding using Fourier expansions like the ones discussed in this paper could perhaps yield a good estimate of the effect of uncorrelated noise on such systems.  Because of the second derivative in this system, the OA ansatz no longer applies, and a generalized or alternative theoretical framework needs to be developed.
The present work is relevant for applications in many disciplines.  For example, examining the synchronization of oscillators of different frequencies might be helpful for examining spike-time reliability in neurons \cite{galanetal08}. One could examine the interplay of antiferromagnetic coupling and noise-induced synchrony using electronic circuits such as those studied experimentally in \cite{williamsetal}, and our original motivation for studying antiferromagnetic synchrony arose from experiments on nanomechanical oscillators \cite{roukes}.




\section*{Acknowledgements} 
YML was funded in part by a grant from KAUST. 
We thank Sherry Chen for early work on antiferromagnetic synchronization in summer 2007 and Mike Cross for his collaboration on that precursor project.  We thank E.~M. Bollt and J. Sun for helpful comments.






\end{document}